\documentclass{PoS}
\usepackage{graphicx}

\usepackage[dvipsnames]{xcolor}
\def\floatcaption#1#2{ \caption{ #2 \ [#1] \label{#1}} }
\def\floatcaption#1#2{ \caption{#2 \label{#1}} }
\def\ttl#1{{\it #1}}
\def\ttl#1{}

\def\bibi{\bibitem}    

\long \def \blockcomment #1\endcomment{}

\def\a{\alpha}
\def\b{\beta}
\def\c{\chi}
\def\d{\delta}
\def\e{\epsilon}                
\def\g{\gamma}

\def\j{\psi}

\def\m{\mu}
\def\n{\nu}

\def\p{\pi}                     
\def\s{\sigma}                  
\def\t{\tau}

\def\D{\Delta}
\def\F{\Phi}

\def\L{\Lambda}

\def\S{\Sigma}

\def\cl{{\cal L}}

\def\sl#1{\rlap{\hbox{$\mskip 1 mu /$}}#1}      
\def\Sl#1{\rlap{\hbox{$\mskip 3 mu /$}}#1}      
\def\ie{\mbox{\it i.e.}}
\def\eg{\mbox{\it e.g.}}

\def\half{{1\over 2}}
\def\tr{{\rm tr}\,}

\def\leqx{\,\raisebox{-1.0ex}{$\stackrel{\textstyle <}{\sim}$}\,}

\def\bj{\overline{\j}}

\def\tcl{\tilde\cl}

\def\svev#1{\left\langle #1\right\rangle}       

\def\ta{\tilde\a}

\def\tc{\tilde{c}}
\def\hf{\hat{f}}

\def\ha{\hat{\a}}
\def\hn{\hat{n}}

\def\otherM{M_{\rm non\mbox{-}NGB}}

\title{Effective field theory for pions and a dilatonic meson}

\ShortTitle{EFT for pions and a dilatonic meson}

\author{\speaker{Maarten Golterman}\\
Department of Physics and Astronomy, San Francisco State University,\\
San Francisco, CA 94132, USA\\
E-mail: \email{maarten@sfsu.edu}}

\author{\speaker{Yigal Shamir}\\
        Raymond and Beverly Sackler School of Physics and Astronomy\\
        Tel-Aviv University, Ramat~Aviv, 69978~Israel\\
        E-mail: \email{shamir@post.tau.ac.il}}

\abstract{
Numerical simulations of QCD-like theories which are chirally broken
but exhibit a very small beta function reveal a flavor-singlet scalar,
or ``dilatonic meson,'' which is much lighter than any other states
except pions.  We develop a systematic low-energy
expansion for the pions and the dilatonic meson, in which the presence
of the latter is attributed to an approximate scale symmetry.
In order to justify the power counting we invoke the Veneziano limit,
in which the number of flavors $N_f$ in the fundamental representation
grows in proportion with the number of colors $N_c$, while the ratio $N_f/N_c$
is kept close to, but below, the critical value where the conformal window
is entered.
}

\FullConference{34st International Symposium on Lattice Field Theory
  LATTICE 2016\\
  24--30 July 2016\\
  Southampton, United Kingdom}

\begin{document}

\section{Introduction}

Asymptotically free gauge theories with relatively few fermion
degrees of freedom exist in a chirally broken and confining phase,
associated with a coupling that grows toward the infrared.
Increasing the number of fermion degrees of freedom can bring the running
of the coupling to a halt.  An infrared-attractive fixed point
(IRFP) appears \cite{IRFP} and the theory exists in an infrared-conformal phase.
The smallest number of flavors where the theory admits an IRFP is
generally referred to as the ``sill'' of the so-called conformal window.

With a number of flavors slightly below the sill, the theory is still
chirally broken
and confining.  But it is different from QCD in being nearly conformal.
More precisely, the beta function is very small near the energy
scale where chiral symmetry breaking sets in.
We say that the theory has a ``walking,'' rather than ``running,'' coupling.

Lattice simulations of walking theories have revealed
the presence of a flavor-singlet scalar meson that can be as light as
the pions over a wide fermion-mass range
(for a recent review, see Ref.~\cite{TD}).  Notable examples
include the $SU(3)$ gauge theory with $N_f=8$ Dirac fermions
in the fundamental representation \cite{LatKMI,LSD} or
with two flavors of sextet fermions \cite{LatHC}.  We stress
that, when dealing with a theory with a very small beta function,
deciding whether the theory is chirally broken and confining, or,
alternatively, infrared conformal, can be very challenging.
Here we will assume that the models mentioned above are indeed
chirally broken in the continuum limit.

Walking theories have features which are attractive for extensions
of the Standard Model that involve a new strong interaction.
The renormalized coupling is changing very slowly with energy scale
even when its value is rather large.  As a result,
one sometimes finds large anomalous dimensions, which, in turn,
can lead to a very large enhancement of the corresponding operator.
This feature is desired when trying to reconcile flavor physics
with experiment (for reviews, see Refs.~\cite{HS,BSM,JG}).  Having a very light
scalar is an added benefit, because, within the context of technicolor-like
theories, it is a natural candidate for the Higgs particle.

Walking theories are also theoretically interesting.  In particular,
it is natural to ask if the presence of the light singlet scalar meson
is somehow connected to the smallness of the beta function.
Indeed, the running of the coupling reflects
the breaking of classical scale invariance by the quantum theory.
When the beta function is small, the quantum breaking of dilatation symmetry
is in some sense also small.
Here we will discuss the construction of a low-energy effective action
for the pions together with the light singlet scalar meson \cite{EFT}.
A consistent low-energy description must account for all the light states,
and must incorporate the scalar meson which can be as light as the pions.
More generally, even if the pions will eventually become lighter
than the scalar meson in the chiral limit, such an effective description
is appropriate whenever the scalar meson is much lighter than
all other states in the theory.  The main challenge facing the
construction is that, in order to build a systematic low-energy
expansion, one has to quantify the violations of dilatation symmetry
in the effective theory, and to be able to relate them to the
microscopic theory in such a way that these violations are controlled
by a small parameter.  The light scalar, or ``dilatonic meson,''
then becomes a pseudo Nambu-Goldstone boson of the
approximate dilatation symmetry.

\section{Building an effective field theory \label{build}}

We start by reviewing the ingredients of standard chiral perturbation theory
(for a review, see Ref.~\cite{MGrev}).
The massless microscopic theory has chiral symmetry, whose spontaneous
breaking gives rise to Nambu-Goldstone bosons, the pions.
When the fermions are given a non-zero mass, the pions become massive, too,
but they remain the lightest asymptotic states as long as the fermion mass
is small enough.

Let us assume that we have $N_f$ Dirac fermions in the fundamental
representation.  This is a complex representation (when $N_c\ge 3$),
and the symmetry breaking pattern is $SU(N_f)_L\times SU(N_f)_R \to SU(N_f)_V$,
where $SU(N_f)_V$ is the diagonal subgroup.  The lagrangian of the
microscopic theory is
\begin{equation}
  \cl^{\rm MIC}(\c) = \frac{1}{4}F^2 + \bj \Sl{D}\j
  + \bj_R\c^\dagger \j_L + \bj_L \c \j_R
\label{Lmic}
\end{equation}
Here $\c$ is an $N_f\times N_f$ matrix-valued spurion,
\ie, an external source field. As usual, $\j_{R,L}=\half(1\pm\g_5)\j$
and $\bj_{R,L}=\half\bj(1\mp\g_5)$.
Under a chiral rotation, the (dynamical) fermion fields
and the (external) spurion field transform according to
\begin{eqnarray}
  \j_{L,R} &\to& g_{L,R}\, \j_{L,R} \ , \qquad
  \bj_{L,R} \ \to\ \bj_{L,R}\, g_{L,R}^\dagger \ , \qquad
  \c \ \to\ g_L \,\c\, g_R^\dagger \ ,
\label{chiralrot}
\end{eqnarray}
where $g_{L,R} \in SU(N_f)_{L,R}$.  The lagrangian $\cl^{\rm MIC}(\c)$
is invariant when we apply the chiral transformation to all the fields
including the spurion field.  The lagrangian is also chirally invariant
when we turn off the external source by setting $\c(x)=0$,
and $\cl^{\rm MIC}(0)$ is recognized as the lagrangian of the massless theory.
But we can also choose to set the chiral source to some
non-zero ``expectation value,'' $\c(x)=m$.
Now $\cl^{\rm MIC}(m)$ is no longer chirally invariant,
and instead, under an infinitesimal chiral transformation we have
$\d\cl^{\rm MIC}(m) = m \d(\bj\j)$, which exhibits the explicit (soft)
breaking of chiral symmetry by the fermion mass term.
We see the dual role of the chiral spurion.  On the one hand,
it encodes the explicit breaking of chiral symmetry coming from
the mass term.  On the other hand, it does so in a manner that
assigns certain chiral transformation properties to the mass matrix itself,
thereby rendering the lagrangian of the {\it massive} theory formally
invariant.  These same transformation properties will next be used to constrain
the structure of the chiral lagrangian.

At the leading order, the lagrangian of the low-energy effective theory is
\begin{equation}
  \cl^{\rm EFT} =
  \frac{f^2}{4}\, \tr(\partial_\m \S^\dagger \partial_\m \S)
   -\frac{f^2 B}{2} \, \tr\Big(\c^\dagger \S + \S^\dagger \c\Big) \ .
\label{LChPT}
\end{equation}
It depends on two low-energy constants (LECs): $f$ and $B$.
The dynamical effective field $\S$ takes values in the coset
$SU(N_f)_L\times SU(N_f)_R / SU(N_f)_V$, which is isomorphic to $SU(N_f)$.
The effective field $\S_{ij}$ is loosely identified with the fermion bilinear
$\tr(\j_{L,i} \bj_{R,j})$, and inherits its transformation properties,
\begin{equation}
  \S \to g_L \,\S\, g_R^\dagger \ .
\label{transS}
\end{equation}
It is easy to check that the chiral lagrangian~(\ref{LChPT}) is invariant
under the combined transformation of Eqs.~(\ref{chiralrot}) and~(\ref{transS}).
Setting $\c(x)=m>0$, it becomes
\begin{equation}
  \cl^{\rm EFT} = - f^2 B m N_f
  + \tr\big( (\partial_\m \p)^2 + 2m B\, \p^2 \big) + O(\p^4) \ ,
\label{treeprop}
\end{equation}
where we have expanded the non-linear field $\S(x)=\exp(2i\p(x)/f)$
around its classical vacuum $\svev{\S}=1$.
We see that at tree level,
the pion mass is given by $M^2=2mB$.  The other LEC, $f$, is
the pion decay constant in the chiral limit (up to normalization conventions),
as can be seen by coupling the effective theory to an external
axial gauge field.

Why does the leading-order chiral lagrangian~(\ref{LChPT}) contain
just two terms?
The chiral lagrangian provides a systematic expansion
in the external momenta and in the fermion mass.
Denoting by $\d$ the small expansion parameter, the power counting is
\begin{equation}
  p^2/\L^2 \ \sim \ m/\L \ \sim \ \d \ .
\label{pcm}
\end{equation}
Here $p^2$ stands for the inner product of any two external momenta.
The reference scale is usually taken to be $\L =4\p f$.
While being a dynamical, infrared scale of the microscopic theory,
$\L$ may be identified with the ultraviolet cutoff of the
chiral lagrangian.  This works because the mass of the pions,
which sets the energy scale probed by the effective lagrangian,
tends to zero in the chiral limit.
At the leading order, we allow for terms of order $\d^1$,
and, after imposing the invariance under chiral symmetry,
this leaves us with just the two operators we have in Eq.~(\ref{LChPT}).

We have seen how the spurion $\c$ communicates information about
the explicit breaking of chiral symmetry between the microscopic
and the effective theories.  More generally, by taking derivatives
with respect to $\c(x)$ and $\c^\dagger(x)$
one defines a set of correlation functions
that can be computed in both theories and compared.
The LECs of the effective theory are fixed order by order
in the chiral expansion~(\ref{pcm}) by requiring that the effective
theory reproduce the correlation functions of the microscopic theory.

\medskip

We now turn our attention to scale transformations, which act
on both the coordinates and the fields.  Given some field $\F(x)$,
its variation under an infinitesimal dilatation is
\begin{equation}
  \d\F = x_\m \partial_\m \F + s\, \F \ ,
\label{diltrans}
\end{equation}
where $s$ is the scaling dimension of $\F$.
In a theory containing gauge and fermion fields (but no elementary scalar
fields) the dilatation current is given by
\begin{equation}
  S_\m = x_\n T_{\m\n} \ ,
\label{SxT}
\end{equation}
where $T_{\m\n}$ is the energy-momentum tensor.
Classically, the lagrangian of the massless theory transforms into
a total derivative under an infinitesimal dilatation,
and the dilatation current is conserved.
Quantum mechanically, the dilatation current is not conserved.  On shell,
its divergence is equal to the trace of the energy-momentum tensor \cite{CDJ}
\begin{equation}
  \partial_\m S_\m  = T_{\m\m} \equiv -T  \ ,
\label{dS}
\end{equation}
where $T = T_{cl} + T_{an}\,,$ and
\begin{equation}
  T_{cl}(m) = m \bj\j \ , \hspace{5ex}
  T_{an}(m) = \, \frac{\b(g^2)}{4g^2}\, F^2 + \g_m\, m\, \bj\j \ .
\label{rentrace}
\end{equation}
All quantities occurring on the right-hand side are
the renormalized ones.  $\b(g^2)$ is the familiar beta function,
while $\g_m=\g_m(g^2)$ is the mass anomalous dimension.
$T_{cl}$ is the classical divergence of the
dilatation current, which vanishes if the fermion mass does.
$T_{an}$ quantifies the quantum breaking of scale symmetry,
reflected primarily in the running of the coupling.

Following the example of chiral perturbation theory,
our first task is to formally recover dilatation invariance of
the microscopic theory.  To this end we introduce a new spurion field
$\s(x)$, which we will call the dilaton.  Unlike the homogeneous
transformation rule~(\ref{diltrans}), the infinitesimal variation of
the dilaton field is
\begin{equation}
  \d\s = x_\m \partial_\m\s + 1 \ .
\label{dildil}
\end{equation}
The inhomogeneous term will play a crucial role below.
The renormalized chiral source transforms like an ordinary field,
with the same anomalous dimension as the renormalized mass,
\begin{equation}
  \d\c = x_\m \partial_\m\c + (1+\g_m) \c \ .
\label{dilchi}
\end{equation}
The lagrangian of the microscopic theory becomes
\begin{equation}
  \cl^{\rm MIC}(\s,\c)
  = \cl^{\rm MIC}(\c) + \s T_{an}(\c) + O(\s^2) \ ,
\label{Lmicsig}
\end{equation}
where $T_{an}(\c)$ is obtained by the replacement $m\to\c(x)$
in Eq.~(\ref{rentrace}).  The classical variation of the lagrangian
is absent thanks to the scale transformation properties
of the chiral source $\c$.  Disregarding total derivatives, the variation
of $\cl^{\rm MIC}(\c)$ is thus $-T_{an}(\c)$, which in turn is cancelled
by the inhomogeneous term in Eq.~(\ref{dildil}) when we vary $\s T_{an}(\c)$.
In order to cancel the terms proportional to $\s$ (as well as to higher
powers of $\s$) in the variation of $\cl^{\rm MIC}(\s,\c)$,
we would need the $O(\s^2)$ terms on the right-hand side of Eq.~(\ref{Lmicsig}).
We will not attempt to derive these higher order terms, because
they do not play any role in the following.

In the case of the chiral lagrangian, we have seen that setting
$\c(x)=0$ reproduces the massless theory, and, hence, exact chiral symmetry.
The same is not true for scale symmetry.  Setting $\c(x)=\s(x)=0$,
the quantum variation of the massless theory becomes $-T_{an}(0)$, namely,
the trace anomaly is $(\b(g^2)/(4g^2))\, F^2$.  The massless quantum theory
is not scale invariant, because the coupling runs.

Moving on to the effective theory, we introduce a new effective field
for the dilatonic meson, denoted $\t(x)$.  Its transformation rule
is similar to that of the external dilaton source,
\begin{equation}
  \d\t = x_\m \partial_\m\t + 1 \ ,
\label{diltau}
\end{equation}
and again contains an inhomogeneous piece.
Both $\s$ and $\t$ are inert under chiral transformations.
As for the non-linear chiral field $\S$, its scaling dimension
must be zero because it is unitary, and its variation under an
infinitesimal dilatation is thus
\begin{equation}
  \d\S = x_\m \partial_\m\S \ .
\label{dilS}
\end{equation}

The next step is to construct the leading-order effective lagrangian.
We are to write down all possible operators that depend on
the effective fields, $\S$ and $\t$, and on the source fields, $\c$ and $\s$,
which are invariant under chiral and scale transformations.
As a first attempt, we follow the same power counting as for the
chiral lagrangian, \ie, we allow for all terms which are of order $\d^1$
according to Eq.~(\ref{pcm}).  The resulting leading-order lagrangian is
\begin{equation}
  \tcl = \tcl_\p + \tcl_\t + \tcl_m + \tcl_d \ ,
\label{LeffL}
\end{equation}
where
\begin{eqnarray}
  \tcl_\p &=& \frac{f_\p^2}{4}\, V_\p(\t-\s)\, e^{2\t} \,
              \tr(\partial_\m \S^\dagger \partial_\m \S) \ ,
\label{LpV}\\
  \tcl_\t &=&
  \frac{f_\t^2}{2}\, V_\t(\t-\s)\, e^{2\t} (\partial_\m \t)^2  \ ,
\label{LtV}\\
  \tcl_m &=& -\frac{f_\p^2 B_\p}{2} \, V_M(\t-\s)\, e^{y\t} \,
  \tr\Big(\c^\dagger \S + \S^\dagger \c\Big) \ ,
\label{LmV}\\
  \tcl_d &=& f_\t^2 B_\t \, V_d(\t-\s)\, e^{4\t} \ .
\label{LdV}
\end{eqnarray}
$\tcl_\p$ and $\tcl_\t$ are the kinetic terms for pions
and for the dilatonic meson, respectively.  $\tcl_m$ is a generalized
chiral mass term, whereas $\tcl_d$ accounts for the self-interactions
of the dilatonic meson.  The presence of a separate set of $f$ and $B$
parameters for the pions and for the dilatonic meson is to be expected.
As we discuss below, the exponent $y$ in Eq.~(\ref{LmV}) compensates
for the dependence of the transformation rule of the renormalized
chiral source on the mass anomalous dimension.

The trouble with this new effective lagrangian is the occurrence
of the potentials $V_\p$, $V_\t$, $V_M$ and $V_d$, each of which
is an arbitrary function of its argument.  The reason why these
potentials are there is that the inhomogeneous terms in the variations
of $\s$ and $\t$ cancel out in the difference $\t-\s$.
As a result, any function $V(\t-\s)$ transforms homogeneously
and has a scaling dimension equal to zero, much like the non-linear field $\S$.
But unlike the $\S$-dependent terms, whose structure is constrained
algebraically both by the unitarity of $\S$ and by the non-abelian nature of
chiral symmetry, the {\it abelian} dilatation symmetry
places no algebraic constraints on the form of the $V(\t-\s)$ potentials.

At this point, our effort seems to have reached a dead end.
The four potentials occurring in the leading-order lagrangian
can be Taylor expanded,
and the expansion coefficients amount to an infinite set of parameters.
If all of them would remain in the leading-order lagrangian,
then we will have lost any predictive power.

To remedy this, we will reexamine the dynamics, seeking a way to extend
the chiral power counting~(\ref{pcm}) to a more powerful one that will impose
a power-counting hierarchy on the Taylor coefficients of these potentials.

\begin{figure}[t]
\begin{center}
\includegraphics*[width=8cm]{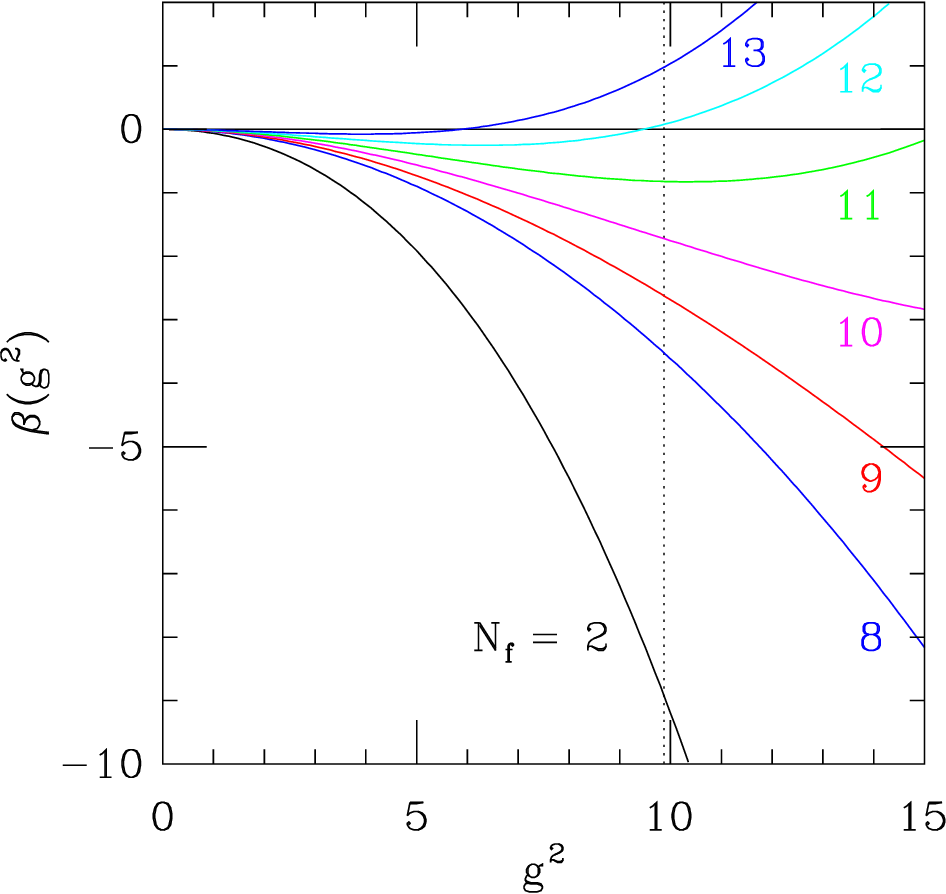}
\end{center}
\begin{quotation}
\floatcaption{betafn}%
{Two-loop beta function of the $SU(3)$ gauge theory with varying
numbers $N_f$ of fundamental-representation flavors.
The dashed vertical line at $g^2=\p^2\simeq 9.87$ marks the critical value
$g_c^2$ of
the coupling where, according to the gap equation, chiral symmetry breaking
takes place in a walking theory.}
\end{quotation}
\vspace*{-6ex}
\end{figure}

\section{A crude model \label{model}}

In this section we consider a crude model for the dynamics of $SU(N_c)$
gauge theories with $N_f$ fermions in the fundamental representation.
As an approximation for the beta function we will consider the familiar
two-loop expression \cite{IRFP},
\begin{equation}
  \frac{\partial g^2}{\partial \log\m}
  = -\frac{b_1}{16\p^2}\, g^4  -\frac{b_2}{(16\p^2)^2}\, g^6 \ .
\label{2loop}
\end{equation}
In Fig.~\ref{betafn} we have plotted the two-loop beta function
for $N_c=3$ and various values of $N_f$.  The $N_f=2$ curve
shows how the beta function looks in a QCD-like theory.
In this case the coefficients $b_1,b_2$ in Eq.~(\ref{2loop}) are both positive,
and the running becomes faster with growing $g$.  As the number
of flavors $N_f$ increases, we reach a range where $b_1>0>b_2$ (for $N_c=3$
this range is given by $8.05 \ \leqx\ N_f < 16.5$).  With $b_1>0$
the theory is still asymptotically free, and the beta function
starts off negative.
But as the coupling grows the screening effect
of the fermions takes over.  The beta function turns back and crosses the axis.
The crossing point $g=g_*$ defines an IRFP.  When $N_f$ is only slightly
above the minimum needed to produce a negative $b_2$,
the value of $g_*$ is very large.
But it decreases monotonically with increasing $N_f$.

As an analytic handle on chiral symmetry breaking
we will use the gap equation.  It predicts that in a walking theory,
chiral symmetry breaking sets in when the coupling reaches
the critical value \cite{HS}
\begin{equation}
  g_c^2 = \frac{4\p^2}{3 C_2} = \p^2 \ ,
\label{gcrit}
\end{equation}
where the last equality is valid for the fundamental representation of $SU(3)$.
Note that $g_c$ 
does not depend on the number of flavors.

We are now ready to determine the ``phase diagram.''
First assume that $N_f$ is small enough that either there is no two-loop IRFP,
or, if it exist, that $g_*(N_f)>g_c$.  As we go down in energy scales,
the coupling $g$ will grow, and chiral symmetry breaking (ultimately
accompanied by confinement) will set in when $g$ reaches $g_c$.
If, on the other hand, $N_f$ is large enough that $g_*(N_f)<g_c$,
the running will come to a halt at the IRFP $g_*$.
The renormalized coupling will never reach $g_c$, and
the infrared physics will be conformal.

Our crude dynamical model predicts that the conformal window
occupies the range $N_f^* \le N_f \le (11/2) N_c$,
where the sill of the conformal window, $N_f^*$,
is the solution of $g_*(N_f^*)=g_c$.
(In general $N_f^*$ is not an integer.  The model
suggests that $N_f^*$ is close to 12 for $N_c = 3$, but whether this is
indeed the case is still under investigation.)
Moreover, the dynamical model reveals an interesting feature
of the chirally broken phase.  As can be seen from Fig.~\ref{betafn},
when $N_f<N_f^*$ and $N_f^*-N_f$ is not too large, the (negative)
beta function at the critical coupling, $\b(g_c^2)$, is
roughly proportional to $N_f-N_f^*$.  This is the hint that will lead us
to the desired power counting.

\section{Power counting \label{showpc}}

According to the model of the previous section,
the beta function at the chiral symmetry breaking scale is $\b(g_c^2)$.
This is a measure of the explicit breaking of dilatation symmetry felt
by the low-energy sector.  As $N_f$ is increased towards the sill of the
conformal window, we expect this explicit breaking to vanish; for $N_f>N_F^*$,
the infrared theory has an emergent conformal symmetry.

Loosely speaking, what this means is that the small parameter controlling
the explicit breaking of dilatation symmetry in the low-energy theory
is $N_f-N_f^*$.  But there is an obvious problem.  $N_f$ takes integer values,
and, unlike the fermion mass, we cannot tune $N_f-N_f^*$ continuously,
nor can we actually reach the critical point $N_f=N_f^*$ since
$N_f^*$ is not an integer.

This problem can be solved for fermions in the fundamental representation
by taking a suitable large-$N$ limit, the Veneziano limit.
We assume that the number of flavors $N_f$ grows in proportion
with the number of colors $N_c$, while the ratio
\begin{equation}
  n_f = N_f/N_c \ ,
\label{nf}
\end{equation}
is held fixed.  Based on the behavior of the two-loop beta function,
we expect that the limit
\begin{equation}
  n_f^* = \lim_{N_c\to\infty} \frac{N_f^*(N_c)}{N_c} \ ,
\label{nfstar}
\end{equation}
will be finite, where now $N_f^*(N_c)$ is an integer:  the actual smallest
number of flavors where the $SU(N_c)$ theory is infrared conformal.
The small parameter we seek for our power counting is $n_f-n_f^*$.
In the Veneziano limit, $n_f$ has effectively become
a continuous parameter, and the Veneziano-limit sill of the conformal window
can be reached by letting $n_f\to n_f^*$ from below.
Of course, we must not forget that the increments we can make in $n_f$
cannot be parametrically smaller than $1/N_c$.  The complete power counting
we need is thus given by (with $N\equiv N_c$)
\begin{equation}
  p^2/\L^2 \ \sim \ m/\L
  \ \sim \ 1/N \ \sim \ |n_f-n_f^*| \ \sim \ \d \ .
\label{pc}
\end{equation}

For any large-$N$ limit, the appropriate coupling is the 't~Hooft coupling,
which we take to be $\ta=g^2 N_c/(16\p^2)$.
Notice that $\b(g^2)/(4g^2)=\b(\ta)/(4\ta)$.
Our central {\it hypothesis} is that at the dynamical scale $\L$
where chiral symmetry breaks spontaneously, the beta function behaves like
\begin{equation}
  \b(\ta(\L)) = O(n_f-n_f^*) + O(1/N) \ .
\label{tbgN}
\end{equation}
As a consequence, $\b(\ta(\L))$ vanishes when the Veneziano limit
followed by the limit $n_f\nearrow n_f^*$ are taken.

We need to spend a moment to explain what $\L$ is.
Let us reexamine Eqs.~(\ref{LpV}) and~(\ref{LtV}).
If we disregard the potentials $V_\p$ and $V_\t$ (the justification for
doing this will be explained shortly),
the pion decay constant in the chiral limit
is $\hf_\p = e^{v_0} f_\p$, where $v_0$ is the expectation value
of the dilatonic meson field in the chiral limit.  Similarly, the decay constant
of the dilatonic meson itself is $\hf_\t = e^{v_0} f_\t$.
Much like $\hf_\p$, the decay constant of the dilatonic meson is defined
by the matrix element of the dilatation current between the vacuum
and a one dilatonic-meson state.  Alternatively, it can be defined
from the matrix element of the energy-momentum tensor between
the same states.  Taking into account the behavior of these matrix elements
in the Veneziano limit, we let
\begin{equation}
  \L \sim \frac{4\p \hf_\p}{\sqrt{N}} \sim \frac{4\p \hf_\t}{N} \ .
\label{LIR}
\end{equation}
Being $O(1)$ in large-$N$ counting, $\L$ is
the characteristic scale for the masses of the lightest {\it non-Goldstone}
mesons, which, in turn, provides the ultraviolet cutoff of
the chiral lagrangian.

How does the power counting~(\ref{pc}) constrain the potentials?
Let us differentiate the lagrangian of the microscopic theory, Eq.~(\ref{Lmicsig}),
with respect to the dilaton source $\s(x)$, and then set the sources to zero.
We obtain
\begin{equation}
  \frac{\partial}{\partial \s(x)}\, \cl^{\rm MIC} \bigg|_{\s=\c=0}
  \ = \ T_{an}(x)\bigg|_{\c=0}
  \ = \ \frac{\b(\ta)}{4\ta}\, F^2(x) \ = \ O(\d) \ ,
\label{dsMIC}
\end{equation}
where the last equality follows from our central assumption~(\ref{tbgN}).
More generally, if we differentiate the partition function $Z^{\rm MIC}$
with respect to the $\s$ field $n$ times, and we are careful to do this
at non-coinciding points, the resulting correlation function will be
parametrically of order $\d^n$.

On the effective field theory side, taking $n$ derivatives  of the lagrangian
with respect to $\s$ probes the $n$-th derivative
of the potentials, $V^{(n)}$.  In terms of the Taylor expansion
\begin{equation}
  V = \sum_{n=0}^\infty \frac{c_n}{n!}\, (\t-\s)^n \ ,
\label{expandV}
\end{equation}
this probes $c_k$ for $k\ge n$.
The idea is to match suitable correlation functions of the microscopic
and the effective theory, setting $\s=0$ (and, if desired, $\c=0$ as well)
in the end.  It takes a detailed study to verify that one can constrain
all the expansion coefficients of the potentials this way \cite{EFT}.
The end result is that the Taylor coefficients
are subject to the power-counting hierarchy
\begin{equation}
  c_n = O(\d^n) \ .
\label{cn}
\end{equation}
The alert reader will have noticed that we must allow for
multiple $\s$ derivatives
at the same spacetime point in the effective theory, but we disallow them
in the microscopic theory.  In fact, this is not a problem, because
the effective theory deals with hadrons, which are not point-like objects;
the effective theory cannot resolve spacetime distances smaller than $1/\L$.

We use this opportunity to draw the attention of the reader to a subtle
point concerning the power-counting proof of Ref.~\cite{EFT}.  While we expect
the hierarchy~(\ref{cn}) to hold for generic (small) values of all of
the expansion parameters~(\ref{pc}), the proof we have given in Ref.~\cite{EFT}
effectively invokes the Veneziano limit, in that it neglects
all the $1/N$ corrections in Eq.~(\ref{tbgN}).  Some other places in Ref.~\cite{EFT}
also tacitly neglect $1/N$ corrections, notably Sec.~4.4, where we discuss
the tree-level theory in the limit $n_f\nearrow n_f^*$.

The final result is that the leading-order lagrangian now consists
of terms of order $\d$ according to the power counting~(\ref{pc}),
with the expansion coefficients of the potentials subject to Eq.~(\ref{cn}).
This allows us to discard $V_\p$, $V_\t$ and $V_M$, because
$\tcl_\p$, $\tcl_\t$ and $\tcl_M$ are already $O(\d)$ without them.
Only in $V_d$ do we need to go to the first non-trivial order in its
expansion.  After setting $\s=0$ and $\c=m$, the leading order lagrangian reads
\begin{equation}
  \cl = \cl_\p + \cl_\t + \cl_m + \cl_d \ ,
\label{Leff}
\end{equation}
where
\begin{eqnarray}
  \cl_\p &=& \frac{f_\p^2}{4}\, e^{2\t} \,
            \tr(\partial_\m \S^\dagger \partial_\m \S) \ ,
\label{Lp}\\
  \cl_\t &=& \frac{f_\t^2}{2}\, e^{2\t} (\partial_\m \t)^2  \ ,
\label{Lt}\\
  \cl_m &=& -\frac{f_\p^2 B_\p}{2} \, e^{y\t} \, m \, \tr(\S + \S^\dagger) \ ,
\label{Lm}\\
  \cl_d &=& f_\t^2 B_\t \, e^{4\t} (c_0 + c_1\t) \ .
\label{Ld}
\end{eqnarray}
It remains to discuss the exponent $y$ in Eq.~(\ref{Lm}).
Assuming that the transition into the conformal window is sufficiently smooth
for $\g_m$, one can show that we need $\g_m=\g_m^*$ in the transformation rule
of the renormalized chiral source, Eq.~(\ref{dilchi}),
where $\g_m^*$ is the IRFP value of the mass anomalous dimension
at the sill of the conformal window.  As a result,
\begin{equation}
  y = 3 - \g_m^* \ .
\label{gammay}
\end{equation}
Present day numerical evidence suggests that $0\le \g_m^* \leqx 1$,
and, therefore, $2\leqx y \le 3$.

\section{Tree level}

In this section we consider the leading-order lagrangian for a given theory
with fixed $N_c$ and $N_f$.  We first discuss the classical vacuum
of the dilatonic meson in the chiral limit.  As follows from Eq.~(\ref{Ld}),
for $m=0$ the dilatonic meson's potential is $U(\t) = e^{4\t} (c_0 + c_1 \t)$
up to a dimensionful constant.
This potential is bounded from below provided that $c_1>0$.
The unique, global minimum
of $U(\t)$ is
\begin{equation}
  v_0 = -1/4 - c_0/c_1 \ .
\label{minV}
\end{equation}
[Like all LECs, the actual value of $c_1$ must be determined by
matching the effective theory to the microscopic theory.
Note that only products such as $c_0 B_\t$ or $c_1 B_\t$
have an invariant meaning, much like $m B_\p$ in the case of
the standard chiral lagrangian.  We use this freedom to assume $B_\t>0$.
Self-consistency of the low-energy description then excludes
a negative value for $c_1$.]

Observe that the classical vacuum would become ill-defined for $c_1=0$.
This has the following interesting interpretation. The potentials $V(\t-\s)$
introduced in Sec.~\ref{build} originate from the explicit breaking of
scale invariance in the massless microscopic theory.
This is true, in particular, for $c_1$, which is the {\it only} LEC
in the leading-order lagrangian coming from the expansion of the potentials
(note that the lagrangian~(\ref{Leff}) becomes scale invariant if
we set $m=c_1=0$).  Thus, the stable classical vacuum of the effective theory
ultimately owes its existence to the running of the coupling
in the microscopic theory.
This should not come as a surprise, because, if the vacuum
has a preferred scale (as opposed to a vacuum with no characteristic scale,
or a continuous manifold of vacua
with a gradually changing characteristic scale),
then the theory cannot have exact scale invariance.

The tree-level mass of the dilatonic meson in the chiral limit is
\begin{equation}
  m_\t^2 = 4 c_1 e^{2v_0} B_\t \ .
\label{mtau}
\end{equation}
If we consider the ratio of the dilatonic meson's mass and decay constant
$\hf_\t = e^{v_0} f_\t$, we get
\begin{equation}
  N^2 m_\t^2/\hf_\t^2 = 4 c_1 N^2 B_\t/f_\t^2 \ ,
\label{mftau}
\end{equation}
in which the dependence on $v_0$ cancels out.
[The role of the factor of $N^2$ on both sides is to undo the large-$N$
dependence of the decay constant of the dilatonic meson,
thereby keeping the ratio finite in the Veneziano limit (compare Eq.~(\ref{LIR})).]
Recall that $c_1=O(\d)$ according to Eq.~(\ref{cn}).
It follows that $m_\t \sim \d^{1/2}$.
This resembles the familiar behavior of the pion mass
in ordinary chiral perturbation theory, $m_\p\sim m^{1/2}$.

We next consider the classical vacuum $v(m)$ for $m>0$.
It is implicitly given by
\begin{equation}
  \frac{f_\p^2 B_\p N_f y m}{f_\t^2 B_\t c_1} = 4 v_1(m) e^{(4-y)v(m)} \ ,
\label{solvesdl}
\end{equation}
where $v_1(m) = v(m)-v_0$.  Generically, $v_1(m)$ is $O(1)$,
because $c_1 \sim m \sim \d$ by the power counting.  One can check
that $v_1(m)>0$ for $m>0$, and that $v(m)$ is a monotonically increasing
function. Using Eq.~(\ref{gammay}),
the tree-level masses of the dilatonic meson and the pion are
\begin{eqnarray}
  m_\t^2 &=& 4c_1 B_\t e^{2v(m)} (1+(1+\g_m^*)v_1(m)) \ ,
\label{mintreetau}\\
  m_\p^2 &=& 2m B_\p e^{(1-\g_m^*)v(m)}
  \ = \ \frac{8 c_1 f_\t^2 B_\t}{y f_\p^2 N_f}\, e^{2v(m)} v_1(m) \ .
\label{mintreepion}
\end{eqnarray}
Both $m_\t$ and $m_\p$ are monotonically increasing with $m$.
Interestingly, the dependence of the tree-level pion mass
on the fermion mass $m$ would reduce to that of
ordinary chiral perturbation theory, if $\g_m^*$ happened to be equal to 1,
which is the favored value according to the gap-equation analysis.
For any other value of $\g_m^*$, Eq.~(\ref{mintreepion}) furnishes us
with a prediction of the low-energy theory that distinguishes it from
ordinary chiral perturbation theory.

\section{Approaching the sill of the conformal window \label{sill}}

In this section we study the tree-level predictions of the effective theory
as the sill of the conformal window is approached.  To avoid
technical complications, we will consider only the chiral limit, $m=0$.
Also, as was done in Ref.~\cite{EFT}, we will take the Veneziano limit,
thereby neglecting the $1/N$ corrections in Eq.~(\ref{tbgN}).

In the Taylor series for the potentials~(\ref{expandV}),
each coefficient $c_n$ can in itself be expanded as a power series
in $n_f-n_f^*$,
\begin{equation}
  c_n = \sum_{k=n}^\infty \tc_{nk} (n_f-n_f^*)^k \ .
\label{expandcn}
\end{equation}
The lower limit of the summation comes from the power-counting
hierarchy~(\ref{cn}) (remember that $n_f-n_f^* \sim \d$).
In particular, the tree-level potential in Eq.~(\ref{Ld}) becomes
\begin{equation}
  V_d(\t) = c_0 + c_1 \t
  = \tc_{00} +  (n_f-n_f^*)(\tc_{01} + \tc_{11} \t) \ .
\label{Vdnf}
\end{equation}
Since $n_f<n_f^*$ for chirally broken theories,
the constraint $c_1>0$ translates into $\tc_{11}<0$.

We may ask what happens if we attempt to apply the low-energy expansion
to a theory that lives {\it inside} the conformal window.
Assuming $n_f>n_f^*$, we see that $c_1=(n_f-n_f^*)\tc_{11}$ becomes negative.
A a result, the classical potential becomes
unbounded from below.  The conclusion is that the effective theory
breaks down inside the conformal window.  This is as it should be,
because there is no spontaneous breaking of chiral symmetry
inside the conformal window.  In this sense, the limit
$n_f\nearrow n_f^*$ is qualitatively different, and more singular,
than the chiral limit $m\to 0$.

Let us next examine the dependence of a few observables on $n_f-n_f^*$.
Since we will be comparing observables belonging to different theories,
we must compare dimensionless quantities.  The dependence on $n_f-n_f^*$
may come directly from $c_1=(n_f-n_f^*)\tc_{11}$, or it can also arise
from the behavior of the classical vacuum $v_0$.  In fact, we already have
one such example, namely, the ratio $N m_\t/\hf_\t$ in the chiral limit,
given in Eq.~(\ref{mftau}).  In this case there is no dependence on $v_0$,
and the dependence on $n_f-n_f^*$ comes only from $c_1$.

Before moving on, it is convenient to use the freedom
to shift the $\t$ field by a constant, $\t \to \t + \D$,
in order to simplify the expression for $v_0$.
Given that $n_f-n_f^*$ is one of the small expansion parameters,
we take $\D$ to be independent of $n_f-n_f^*$ so as not to obscure
the power counting.
Substituting in Eq.~(\ref{Vdnf}) we see that the shift has the effect of changing
$\tc_{01} \to \tc_{01} + \tc_{11} \D$, while $\tc_{00}$ and $\tc_{11}$
are unchanged.  We will use this freedom to set $\tc_{01}=0$.
(The remaining dependence of the lagrangian~(\ref{Leff}) on the shift $\D$
is absorbed into redefinitions of the $f$'s and $B$'s.)
The classical vacuum of the $m=0$ theory thus becomes
(compare Eq.~(\ref{minV}))
\begin{equation}
  v_0 = -1/4 -\tc_{00}/(\tc_{11}(n_f-n_f^*)) \ .
\label{minVnf}
\end{equation}
We comment in passing that the dependence of the physical decay constants,
$\hf_\p = e^{v_0} f_\p$ and $\hf_\t = e^{v_0} f_\t$, on $v_0$ suggests
that we should have $v_0\to -\infty$ for $n_f\nearrow n_f^*$,
which in turn requires $\tc_{00}>0$.  Appealing as this may be, however,
we have not been able to prove this assertion, basically because it involves
the comparison of  dimensionful quantities of different theories.

As our second example we consider the fermion condensate,
measured in units of $\hf_\p$.  We find
\begin{equation}
  \frac{\svev{\bj\j}}{\hf_\p^3}
  = -
  \frac{B_\p N_f}{f_\p}\, e^{-\g_m^* v_0} \ ,
\label{condscale}
\end{equation}
where $v_0$ is now given by Eq.~(\ref{minVnf}), and where we have used that
the tree-level condensate is
\begin{equation}
  \svev{\bj\j} = -f_\p^2 B_\p N_f\, e^{yv_0} \ .
\label{eftbjj}
\end{equation}
Assuming that $\tc_{00}>0$ (and that $\g_m^*>0$ as well),
Eq.~(\ref{condscale}) predicts an enhancement of the fermion condensate
for $n_f\nearrow n_f^*$, which, apart from the familiar dependence
on the mass anomalous dimension, depends also on the LECs $\tc_{00}$
and $\tc_{11}$ through Eq.~(\ref{minVnf}).

\medskip

The low-energy effective theory provides us with a quantitative description
of the (pseudo) Nambu-Goldstone sector in the chirally broken phase.
But it does not give us any access to physics inside the conformal window,
nor to the dynamics of a chirally broken theory at any energy scale
which is comparable to or larger than $\L$.
We may gain some qualitative understanding of the transition into
the conformal window by using the dynamical model of Sec.~\ref{model}.
This consists of using the two-loop beta function,
combined with the prediction of the gap equation for the critical coupling
that triggers chiral symmetry breaking.  Here we add a new element,
namely, we will use this dynamical model in the Veneziano limit,
where, in terms of the 't~Hooft coupling introduced in Sec.~\ref{showpc},
the critical coupling is $\ta=1/6$ (for fermions in the fundamental
representation).

In the Veneziano limit, one can express the two-loop beta function as
\begin{equation}
  \b(\ta) = -\left( \frac{1}{6}-\ha \right)^2
  \left( \ha + \frac{\hn}{3} \left( \frac{25}{6} - 13\ha \right) \right) \ ,
\label{betasill}
\end{equation}
where we wrote $\ta = 1/6-\ha = \ta_c-\ha$ and $n_f = 4-\hn$.
At the chiral symmetry breaking scale $\ta(\L)=\ta_c$,
which corresponds to $\ha=0$.  The beta function then
satisfies $\b(\ta_c)\propto\hn$.
It follows that the sill of the conformal window is at $n_f^* = 4$,
and that the conformal window is $4<n_f<11/2$.
(For $n_f>11/2$ asymptotic freedom is lost.)

We next introduce a new reference scale denoted $\L_{nc}$,
where the subscript ``$nc$'' stands for ``nearly-conformal.''
It is defined in the massless theory by the condition that
\begin{equation}
  \b(\ta(\L_{nc})) = -\e_0 \ ,
\label{Lnc}
\end{equation}
for some fiducial value $0<\e_0\ll 1$.
Eq.~(\ref{Lnc}) is supplemented by the additional
instruction that $\L_{nc}$ is to be found by starting in the deep infrared,
and then increasing the scale till Eq.~(\ref{Lnc}) is satisfied.
[This additional instruction is needed to avoid the second occurrence
of $\b(\ta) = -\e_0$ in the vicinity of the gaussian fixed point,
as is visible, for example, in the $N_f=12$ or $N_f=13$ curves in Fig.~1.]

Because it relies on the beta function, the criterion~(\ref{Lnc})
make sense only if its solution $\L_{nc}$ is large compared to any
dynamical infrared scale that may be induced in the massless theory.
The scale $\L_{nc}$ thus always exists for theories inside the conformal window,
where no dynamical infrared scale is generated.
In the chirally broken phase, our dynamical model predicts that
$\L_{nc}$ exists provided that $n_f$ is close enough to $n_f^*$, so that
at the critical coupling, $|\b(\ta_c)|<\e_0$.
Moreover, because $\b(\ta_c)$ tends to zero
when $n_f$ tends to $n_f^*$, it follows that the ratio
$\L/\L_{nc}$ also tends to zero in this limit.

Let us now distinguish three regions for the fermion mass:
\begin{eqnarray}
  {\rm I}:\ \L \ll m \ll \L_{nc} \ ,
\qquad
  {\rm II}:\ m \sim \L \ ,
\qquad
  {\rm III}:\ m \ll \L \ .
\nonumber
\end{eqnarray}
Region III is where the low-energy expansion is valid.
The theory has both approximate chiral symmetry and approximate
dilatation symmetry, both of which are spontaneously broken.

In Region I, chiral symmetry and dilatation symmetry are both explicitly
broken by the fermion mass, but this breaking is soft.
Because of the smallness of the beta function,
what we expect to see in Region I is the characteristic behavior of
a {\it mass-perturbed conformal system}.  This implies that the masses
of all mesons behave like (see \eg\ Ref.~\cite{DDZ})
\begin{equation}
  M \sim \L \left(m/\L\right)^{\frac{1}{1+\g_m^*}} \ .
\label{Mscale}
\end{equation}

The transition between the conformal and chirally broken behavior
occurs in Region II.  Once $m$ goes below $\L$,
we enter the chiral regime.  The masses of all non-Goldstone mesons
freeze out at $\otherM\sim\L$, while the masses of the pseudo
Nambu-Goldstone mesons behave like
\begin{equation}
  M^2_{\rm pNGB} \ = \ \left[
    O(n_f-n_f^*) + O(m/\L)\right] \L^2 \ \ll \ \L^2
  \ \sim \ \otherM^2 \ .
\label{M2L}
\end{equation}
We see that as $n_f$ tends to $n_f^*$ from below, the masses of
{\it all} mesons in the massless theory tend to zero,
if measured in units of $\L_{nc}$.
But the masses of the pseudo Nambu-Goldstone mesons vanish faster;
the smallness of the ratio $M_{\rm pNGB}/\otherM$ is what allows for the
existence of a systematic low-energy description.

Notice that in order to stay in the chiral regime when $n_f$ gets closer
to $n_f^*$ we must keep decreasing $m$.  This is because we must maintain
$m/\L\ll 1$, and $\L/\L_{nc}$ vanishes at the conformal sill.
It is also useful to consider what happens if we hold $m$ fixed
in units of $\L_{nc}$.  Regardless of whether $n_f$ is smaller or larger
than $n_f^*$, all theories where $|n_f-n_f^*|\ll 1$ then have a wide region
where the theory exhibits the typical behavior
of a mass-perturbed conformal system.
The difference between $n_f>n_f^*$ and $n_f<n_f^*$ is that in the former case,
the mass-perturbed conformal behavior exists for any $m\ll \L_{nc}$, regardless
of how small $m$ is. By contrast, for $n_f<n_f^*$ this behavior exists only
in Region I: $\L \ll m \ll \L_{nc}$, which is bounded from below.
As $n_f$ approaches the sill $n_f^*$, the range of fermion mass where
the theory exhibits a mass-perturbed conformal behavior keeps expanding
because $\L/\L_{nc}$ gets smaller, until eventually at $n_f=n_f^*$
we have $\L/\L_{nc}\to 0$, and the chirally broken behavior
is completely lost.
The physical picture that emerges is that, if we always use $\L_{nc}$
as the reference scale, and the fermion mass is kept at some
fixed value in units of $\L_{nc}$, then the physical spectrum
will vary continuously as we dial $n_f$ upwards, across $n_f^*$
and into the conformal window.  In this sense,
the transition into the conformal window is smooth.

\vspace{2ex}
\noindent {\bf Acknowledgments}
\vspace{1ex}

We thank David B. Kaplan for raising questions about the behavior
of the theory at the transition to the conformal phase.
This material is based upon work supported by the U.S. Department of
Energy, Office of Science, Office of High Energy Physics, under Award
Number DE-FG03-92ER40711.
YS is supported by the Israel Science Foundation
under grant no.~449/13.

\vspace{0ex}


\begin{thebibliography}{99}

\bibi{IRFP}
  W.~E.~Caswell,
  \ttl{Asymptotic behavior of nonabelian gauge theories to two loop order,}
  Phys.\ Rev.\ Lett.\  {\bf 33}, 244 (1974).
  T.~Banks and A.~Zaks,
  \ttl{On The Phase Structure Of Vector-Like Gauge Theories
    With Massless Fermions,}
  Nucl.\ Phys.\  B {\bf 196}, 189 (1982).

\bibi{TD}
  T.~DeGrand,
  \ttl{Lattice tests of beyond Standard Model dynamics,}
  Rev.\ Mod.\ Phys.\  {\bf 88}, 015001 (2016)
  [arXiv:1510.05018 [hep-ph]].

\bibi{LatKMI}
  Y.~Aoki {\it et al.} [LatKMI Collaboration],
  \ttl{Light composite scalar in eight-flavor QCD on the lattice,}
  Phys.\ Rev.\ D {\bf 89}, 111502 (2014)
  [arXiv:1403.5000 [hep-lat]].

\bibi{LSD}
  T.~Appelquist {\it et al.},
  \ttl{Strongly interacting dynamics and the search for new physics
    at the LHC,}
  Phys.\ Rev.\ D {\bf 93}, no. 11, 114514 (2016)
  [arXiv:1601.04027 [hep-lat]].

\bibi{LatHC}
  Z.~Fodor, K.~Holland, J.~Kuti, S.~Mondal, D.~Nogradi and C.~H.~Wong,
  \ttl{Status of a minimal composite Higgs theory,}
  PoS LATTICE {\bf 2015}, 219 (2016)
  [arXiv:1605.08750 [hep-lat]].

\bibi{HS}
  C.~T.~Hill and E.~H.~Simmons,
  \ttl{Strong dynamics and electroweak symmetry breaking,}
  Phys.\ Rept.\  {\bf 381}, 235 (2003)
  Erratum: [Phys.\ Rept.\  {\bf 390}, 553 (2004)]
  [hep-ph/0203079], and references therein.

\bibi{BSM}
  R.~Contino,
  \ttl{The Higgs as a composite Nambu--Goldstone boson,}
  arXiv:1005.4269 [hep-ph].
   B.~Bellazzini, C.~Cs\'aki and J.~Serra,
   \ttl{Composite Higgses,}
   Eur.\ Phys.\ J.\ C {\bf 74}, 2766 (2014)
   [arXiv:1401.2457 [hep-ph]].
   G.~Panico and A.~Wulzer,
   \ttl{The composite Nambu--Goldstone Higgs,}
   Lect.\ Notes Phys.\  {\bf 913}, 1 (2016)
   [arXiv:1506.01961 [hep-ph]].

\bibi{JG}
  J.~Giedt,
  \ttl{Anomalous dimensions on the lattice,}
  Int.\ J.\ Mod.\ Phys.\ A {\bf 31}, no. 10, 1630011 (2016)
  [arXiv:1512.09330 [hep-lat]].

\bibi{EFT}
  M.~Golterman and Y.~Shamir,
  \ttl{Low-energy effective action for pions and a dilatonic meson,}
  Phys.\ Rev.\ D {\bf 94}, no. 5, 054502 (2016)
  [arXiv:1603.04575 [hep-ph]].

\bibi{MGrev}
  M.~Golterman,
  \ttl{Applications of chiral perturbation theory to lattice QCD,}
  arXiv:0912.4042 [hep-lat].

\bibi{CDJ}
  J.~C.~Collins, A.~Duncan and S.~D.~Joglekar,
  \ttl{Trace and Dilatation Anomalies in Gauge Theories,}
  Phys.\ Rev.\ D {\bf 16}, 438 (1977).

\bibi{DDZ}
  L.~Del Debbio and R.~Zwicky,
  \ttl{Hyperscaling relations in mass-deformed conformal gauge theories,}
  Phys.\ Rev.\ D {\bf 82}, 014502 (2010)
  [arXiv:1005.2371 [hep-ph]].

\end{thebibliography}
\end{document}